\documentclass[sigconf]{acmart}

\acmDOI{10.1145/1122445.1122456}

\acmISBN{978-1-4503-7163-6}
\acmConference[ICFNDS]{ACM International Conference on Future Networks and Distributed Systems}{July 2019}{Paris, France} \acmYear{2019} \copyrightyear{2019} 

\begin{document}

\title{A Pub/Sub SDN-Integrated Framework for IoT Traffic Orchestration}

\author{Pedro F. Moraes}
\email{pmoraes_@hotmail.com}
\affiliation{%
  \institution{Salvador University - UNIFACS}
  \streetaddress{Rua Dr. Jose Peroba 251}
  \city{Salvador}
    \country{Brazil}
  \postcode{41770-235}
}

\author{Joberto S. B. Martins}
\email{joberto.martins@gmail.com}
\affiliation{%
  \institution{Salvador University - UNIFACS}
  \city{Salvador}
  \country{Brazil}
}

\renewcommand{\shortauthors}{Moraes, Pedro F., et al.}

\begin{abstract}
 The Internet of Things (IoT) is advancing and the adoption of internet-connected devices in everyday use is constantly growing. This increase not only affects the traffic from other sources in the network, but also the communication quality requirements, like Quality of Service (QoS), for the IoT devices and applications. With the rise of dynamic network management and dynamic network programming technologies like Software-Defined Networking (SDN), traffic management and communication quality requirements can be tailored to fit niche use cases and characteristics. We propose a publish/subscribe QoS-aware framework (PSIoT-SDN) that orchestrates IoT traffic and mediates the allocation of network resources between IoT data aggregators and pub/sub consumers. The PSIoT framework allows edge-level QoS control using the features of publish/ subscribe orchestrator at IoT aggregators and, in addition, allows network-level QoS control by incorporating SDN features coupled with a bandwidth allocation model for network-wide IoT traffic management. The integration of the framework with SDN allows it to dynamically react to bandwidth sharing enabled by the SDN controller, resulting in better bandwidth distribution and higher link utilization for IoT traffic.
\end{abstract}




\maketitle

\section{Introduction}
{T}{he} Internet of Things (IoT) is considered an important trend in the future Internet  \cite{These-Elie} \cite{borgia2014}. Its ability to connect common objects to each other and through the internet has many different applications. The growing number of these connected devices brings with them many advantages to businesses, consumers, cities and plays a role in emerging technology.

Most of IoT applications have a large amount of heterogeneous devices in the form of sensors and actuators, with differences in processing, storage, power and functionality. These differences then generate a large amount of heterogeneous traffic that leads to complex issues in quality of service (QoS), resource allocation and network configuration.

Current network management already deal with the issues that an overload in traffic creates, and there are already solutions for network management aimed at controlling network resources and traffic priority \cite{baddeley_evolving_2018}. These capabilities are typically  referred to as a network's Quality of Service (QoS). The purpose of QoS is to control the use of network resources more efficiently and ensure the required levels of service quality, using network characteristics such as bandwidth, latency, jitter and reliability.

Additionally, applications on the network can also add specialized QoS capabilities for application SLA guarantees or cost control. The way applications use data, and their characteristics can better define QoS models, such as in video streaming where buffering and image quality are modified according to user behavior and network quality.

Cloud Computing \cite{cloud} and recently Fog Computing \cite{fog} attempt to mitigate the impact of massive IoT traffic has on networks and devices and data processing. Using Fog Computing operating on the edges of the network allows to minimize the load on the whole network by serving already preprocessed and aggregated IoT data. This allows applications to leverage the geographical distribution of IoT data-points to keep the traffic flow closer to the consumers.

IoT applications can benefit from these mechanisms to create more appropriate QoS management. However, with IoT applications and devices spreading geographically and increasing in quantity, data aggregation and preprocessing in the Fog isn't enough to compensate the large increase of IoT traffic in the network. Purely decreasing IoT traffic via aggregation and preprocessing also doesn't automatically enhance the QoS for massive IoT data.

Using IoT data and traffic characteristics is important in order to create QoS strategies that better serve IoT applications. Knowing how IoT devices transmit data and how applications consume that data makes it easier to properly apply QoS in a meaningful way.

The PSIoT-Orch framework is a QoS-aware framework \cite{moraes_publish/subscribe_2018-2} for managing IoT traffic aggregated into Fog-like IoT gateways along the network edge. In this solution aggregators and a orchestrator allow for IoT QoS management at the network edge enabled by network-wide specifications, application domain and IoT characteristics. The PSIoT-SDN framework expands on the usual orchestrator's edge-level QoS control operation by using SDN network programmability coupled with a bandwidth allocation model (BAM). We evaluate the enhancements in QoS when integrating the framework with the SDN network manager and how it positively affects IoT traffic QoS over the network. As such, contribution of the PSIoT-SDN framework will be to integrate a QoS approach on network edge with a network bandwidth sharing strategy based on BAM models that is SDN-controlled.

In this article we'll describe the PSIoT-SDN framework and how it proposes to manage IoT traffic, as well as how it combines with SDN to offer enhanced QoS capabilities to IoT traffic. We present in section \ref{sssec:related} previous work in that handle QoS and traffic management. In section \ref{sssec:iotarch} we explore proposed IoT-oriented architectures, considering both data processing and QoS in an SDN enabled network. Section \ref{sssec:frameworkintro} overviews PSIoT-SDN framework components including its basic orchestrator for IoT traffic management at network edge-level (aggregators) and the SDN controller for network-wide QoS management. In Section \ref{sssec:evaluation} we present a proof of concept for the PSIoT-SDN with a SDN network controller coupled with a bandwidth allocation model (BAM) and evaluate how it affects the traffic in the network. Finally, section \ref{sssec:conclusion} concludes with an overview of the result and what was presented.

\section{Related Work} \label{sssec:related}

IoT applications and devices are varied in communication protocols, hardware and characteristics. This has generated a large amount of IoT-focused works in the literature ranging from hardware and network protocols to applications and traffic protocols and management \cite{bera_software-defined_2017-1}.

A comprehensive survey in \cite{10.1109/COMST.2015.2444095} reviews the enabling technologies, protocols and applications for  IoT, presenting elements required for devices to deliver IoT functionality. The work in \cite{10.1016/j.comcom.2014.09.008} presents the IoT key features and driver technologies, detailing IoT into application domains like industry, smart city and health.


Considering IoT traffic in wireless networks, \cite{10.1109/JIOT.2017.2785219}, \cite{10.1109/IC3I.2016.7917982}, \cite{10.1109/CEEC.2017.8101604} propose traffic management solutions that focus on wireless sensor networks (WSN) and wireless cellphone networks. The research focus is mainly in considering the constraints of low power and computation on IoT devices and wireless networks.


The research categorized in the IoT-SDN integrated approach are mainly focused on methods to manage the IoT traffic being generated by IoT devices after it has entered the network. These are mostly about using the SDN dynamic programming capability and Cloud-related technology to offer better IoT traffic control towards mobility, security, spectrum and service management  \cite{tayyaba_software_2017} \cite{wu_ubiflow_2015} \cite{boussard_software-defined_2015-1} \cite{jararweh_sdiot_2015}.

Using the Fog for edge data processing and aggregation, \cite{10.1109/MNET.2017.1700271} and \cite{10.1109/CloudCom.2017.62} present platforms for orchestrating Fog workload. The research in \cite{ 10.1109/RNDM.2017.8093036} extends Fog computing to improve network resilience in Fog nodes and \cite{10.1109/ACCESS.2017.2778504} does an extensive survey on edge computing for the IoT.

The research in \cite{10.1109/CC.2018.8290809} presents a publish/subscribe communication platform for IoT services. It is based in SDN and directly uses the network's SDN management, and knowledge on the network's routing and topology, to control the QoS of topic subscriptions traffic.

\section{IoT Architecture, Data Processing and SDN Integration}  \label{sssec:iotarch}

The Internet of Things is growing more and more as users and applications arise with IoT connected devices. The IoT enables everyday objects to be identified and to communicate with each other and other applications over the internet. This flexibility also means that the IoT is extremely heterogeneous, in both data and network usage.

As the IoT expands, effort has been made to find common traits and characteristics of the IoT connected devices as well as architectures for IoT data processing and network communication.

\subsection{IoT Architecture}

The heterogeneous nature of IoT devices, data and overlapping application domains makes it difficult to enable interoperability between IoT devices and applications, despite IoT traffic characteristics and requirements being well defined.

In \cite{etsi}, the ETSI Technical Committee for Machine-to-Machine Communications (ETSI TC M2M) proposed an IP-based architecture leveraging existing technologies. The architecture was divided into three domains: 

\begin{itemize}
    \item The Application Domain, where client and M2M applications lie;
    \item The Network Domain, the network between applications and device gateways; and
    \item The Device \& Gateway domain, the location the devices and/or gateways reside.
\end{itemize} 

The definition of these domains are useful in organizing and designing IoT solutions, as it provides a clearer picture on how IoT devices and data happen in networks and the internet. This clearer distinction can allow for tailored  solutions, catering to each domain specifically and allow a better integration between architectures and IoT solutions, like data processing or IoT QoS.

\subsection{IoT Data Processing}

The growing number of IoT devices generates an also growing amount of data being transmitted through the internet and wireless networks. This amount of traffic can burden the network and cause IoT services that rely on timely communication to malfunction. Aggregating and preprocessing IoT data can alleviate the burden on the network. 

This process can happen at two points: in-network (edge and network devices) and at the cloud level \cite{10.3390/s131115582} \cite{1506.09118}. Cloud IoT data processing aims to cut down on IoT traffic to IoT data consumers by processing device data into aggregated data of interest to many consumers.

Further data processing also happens in local IoT device networks, such as wireless, and wireless mesh networks. This occurs because of the energy and resource constraints on these devices, where neighbouring data that is usually highly redundant is aggregated, and reduced in size, thus preserving energy and resources.

Data aggregation  is  responsible  for  increasing  the  network  lifetime  and  reducing  the  energy consumption \cite{tripathi2014survey}. A comparison of data aggregation techniques is presented in \cite{10.1109/WiSPNET.2016.7566346}, and \cite{10.1016/j.jnca.2017.08.006} presents a literature survey on data aggregation mechanisms.

Cloud aggregation and data processing don't fully supply the IoT need for low latency, leading to the search of some derived solutions. Considering this problem, Cisco introduced in \cite{bonomi2012fog} the concept of Fog Computing to enable applications on billions of connected devices already connected in the Internet of Things to run directly in the edge of the network, providing mobility and geo-distribution support, as well as managing location and latency.

A hybrid approach to data aggregation exists, where Fog nodes act as preprocessors that aggregate IoT data in the network edge before forwarding it into the Cloud, presented in \cite{alturki2017hybrid}.

\begin{figure}[!htb]
\includegraphics[width=\columnwidth]{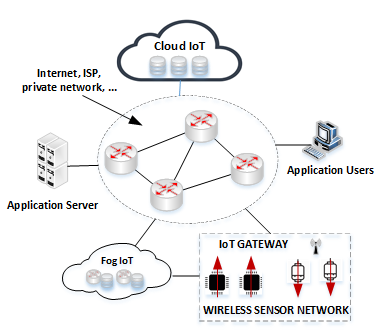}
\caption{IoT architectural basic elements and data processing}
\label{IoTProcessing}
\end{figure}

Considering these data aggregation methods, and the ETSI IoT domain model, Figure \ref{IoTProcessing} shows how the different IoT data aggregation are located in the network. It shows how IoT data flows from devices in the M2M Domain, is aggregated and preprocessed in the Fog and forwarded to the Cloud. This stream of aggregation and processing can cut back on the load that massive IoT traffic can cause in the network.

\subsection{IoT Network Architecture with SDN}

With more complex networks, traditional management paradigms must be reworked and changed. There is a need for network management approaches that are scalable and are able to work well with the heterogeneous, complex, and large-scale nature of IoT.

The great number of heterogeneous IoT devices and data require new technology and management patterns to better serve IoT traffic QoS needs. SDN is a well-matched network programming paradigm suitable for IoT management, that separates the control and data planes. SDN provides high-level abstraction and virtualized network functions that make management simpler \cite{ valdivieso2014sdn}. 

SDN-based IoT management allows for better control over traffic routing and network configuration, allowing the network to more easily react to network traffic needs. The dynamic configuration capabilities of SDN networks are a great match to the heterogeneity of IoT traffic, allowing the network to better route IoT traffic and manage its impact on the whole network.

\begin{figure}[htbp]
\centering
\includegraphics[width=\columnwidth]{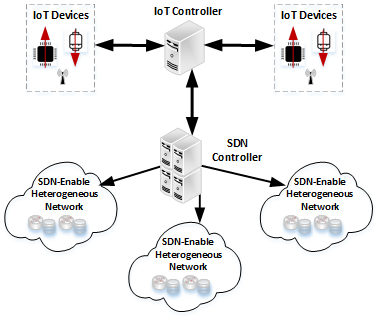}
\caption{IoT architecture with SDN: a high-level view }
\label{sdn-iot}
\end{figure}

Integrating IoT frameworks with SDN gives greater control over IoT traffic. Figure \ref{sdn-iot} shows a high-level view of a a typical architecture for IoT-SDN integration.

The SDN-IoT integration brings several significant benefits for IoT traffic:
\begin{enumerate}
\item	Intelligent traffic routing and better network resources use.
\item	Simplified information acquisition facilitating information analysis, decision making and network configuration actions.
\item   Virtualization, whenever required, may be easily achieved and deployed using common SDN virtualization tools like hypervisors \cite{kreutz2015software}.
\item	Visibility of network resources and access management based on user, group, device, and application.
\item	Intelligent algorithms to build effective traffic pattern analyzers.
\end{enumerate}

These benefits result in IoT networks with integrated SDN capabilities becoming more agile, scalable and based on demand.

\section{The PSIoT Framework Basics} \label{sssec:frameworkintro}

As described in \cite{moraes_publish/subscribe_2018-2}, the PSIoT-Orch is a IoT traffic management framework that combines the Cloud, Fog and Pub/Sub IoT scenarios to enable Pub/Sub-based IoT data transfers and QoS requirements enforcement based on IoT data sets. 

The PSIoT-Orch framework was created to manage massive traffic generated by the
ever growing number of IoT devices. Its goal is to use Publish/ Subscribe to allow IoT data transfer among producers and consumers and, concomitantly, to handle network resources efficiently according to IoT QoS traffic requirements at edge level. It adopts a static priority allocation for IoT traffic data classes at aggregators  \cite{moraes_publish/subscribe_2018-2}.

PSIoT-Orch manages the IoT network resources by leveraging its control on IoT traffic transmission through the network. This management is specially tailored to IoT data and traffic characteristics, enabling better QoS than data-agnostic QoS rules in the network.

Maintaining IoT data aggregators at the edges of the network enables the framework to have more precise knowledge on IoT data, allows pre-processing and all Fog-related benefits while enabling traffic shaping and prioritizing based on the higher priority of IoT data flows.

\subsection{The PSIoT-Orch architecture}
The PSIoT-Orch framework has 3 main components:
\begin{itemize}
    \item A traffic orchestrator;
    \item Fog-like IoT gateway data aggregators (IoTGW-Ag) acting as Pub/Sub publishers; and
    \item Pub/Sub clients, be they end-user applications or Cloud processing centers.
\end{itemize}

\begin{figure}[!htb]
\includegraphics[width=\columnwidth]{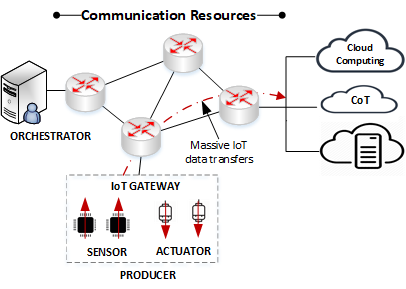}
\caption{PSIoT-Orch framework basic components}
\label{Framework-Components}
\end{figure}

 Figure \ref{Framework-Components} shows how the components are distributed in the network. The communication happens by trading messages through the network, with a centralized orchestrator and multiple clients and producers.

The IoTGW-Ag are Fog-like nodes that act as IoT data aggregators and Pub/Sub producers. These nodes can take other IoT-related responsibilities such as data pre-processing, backup or caching. Each IoTGW-Ag connects to the PSIoT orchestrator, that manages the traffic transmission from the many aggregators in the network.

\begin{figure}[!htb]
\includegraphics[width=\columnwidth]{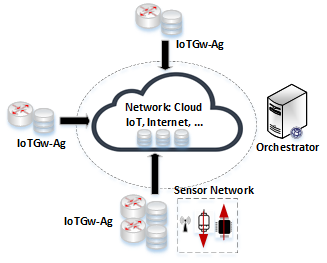}
\caption{PSIoT-Orch architectural components: Orchestrator and IoT Aggregators.}
\label{img:psiot}
\end{figure}

The orchestrator can leverage its control on each IoTGW-Ag in the network to manage the flow of IoT traffic. Figure \ref{img:psiot} illustrates how the framework is positioned in the network and how it can leverage the flow of IoT data by managing the traffic flow constraints at each IoTGW-Ag in the network, according to IoT data characteristics.

The orchestrator has the knowledge of each IoTGW-Ag's Pub/Sub subscriptions, as well as the QoS levels required for each topic subscription. This allows it to decide the transmission rates of IoT data from each aggregator in the network, so as to maximize the throughput of higher level QoS subscriptions \cite{moraes_publish/subscribe_2018-2}.

The PSIoT-Orch framework relies on the time-sensitivity of IoT traffic to determine its QoS levels. While managing aggregators transmission rates doesn't increase the overall throughput of IoT traffic, it does do so to higher priority IoT topic-based subscriptions. Higher priority IoT data can flow faster in comparison to overall IoT traffic load on the network.

In summary, PSIoT-Orch manages QoS at the aggregators on behalf of IoT traffic by managing the traffic in queues with the orchestrator (Figure \ref{img:psiot}). In the PSIoT-Orch architecture QoS level assurance is only suported at edge-level at the aggregators. This approach is suitable for any network infrastructure, including the Internet, where no control exists over the network resources being used.

\section{The PSIoT-SDN architecture}


The PSIoT-SDN framework introduces a network-level QoS control in the PSIoT framework. The target now is to achieve both edge-level and network-level QoS management by leveraging the QoS mangement capabilities at aggregattors provided by the PSIoT orchestrator and to introduce a SDN-based network links bandwidth allocation strategy. For achieving that, the SDN management is coupled with a bandwidth allocation approach for network links resource allocation.  

\begin{figure}[htbp]
\centering
\includegraphics[width=\columnwidth]{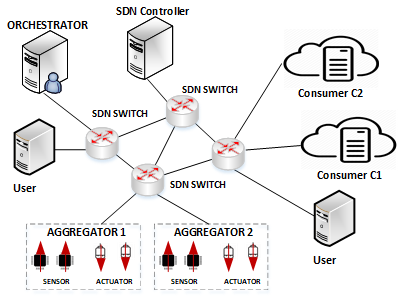}
\caption{ PSIoT-SDN integrated framework}
\label{PSIoT-SDN}
\end{figure}

The components of the PSIoT-SDN architecture are shown in Figure \ref{PSIoT-SDN}. In effect, an SDN controller is introduced in the architecture to allow IoT flows programmability at the switches deployed in the network.

In the PSIoT-SDN architecture it is assumed that, firstly, the IoT traffic is heterogeneous in relation to its QoS requirements and is present in multiple network locations. From these highly distributed locations IoT traffic has to travel from IoT sensors to the aggregators and, from there, using a controlled network to its consumers. These factors make it so that finding a common hierarchy of QoS needs for IoT traffic harder and usually leads to tailoring QoS requirements according to local device and network resources capability.

The PSIoT-SDN framework is built to work on top of a controlled network that it can program directly. It is assumed that the framework has direct control over network link resources or routing decisions using the SDN deployed interface. This makes it so that the IoT QoS management has effectively two components:
\begin{itemize}
    \item An IoT QoS traffic management, enabled by the control of all IoT data producers at the IoTGW-Ag queues; and
    \item A link resource management by adjusting bandwidth between network switches all over the path between aggregators and consumers using SDN. 
\end{itemize}

SDN enables dynamic network configuration, routing and QoS in networks, and can be regarded as an ally when dealing with IoT traffic. Integrating SDN enabled network management into IoT systems can help to fine tune and fulfill IoT QoS requirements.

Integrating SDN network management within the PSIoT-SDN framework enables enhanced QoS with better guarantees on packet QoS due to the fact that link bandwidth will efficiently allocated among network users.

The next section presents the operation flow of a network running the PSIoT-SDN framework with SDN network management and introduces the bandwidth allocation strategy for IoT flows and other traffic in network links

 
\subsection{PSIoT-SDN link bandwidth management and operation flow}

\begin{figure}[htbp]
\centering
\includegraphics[width=\columnwidth]{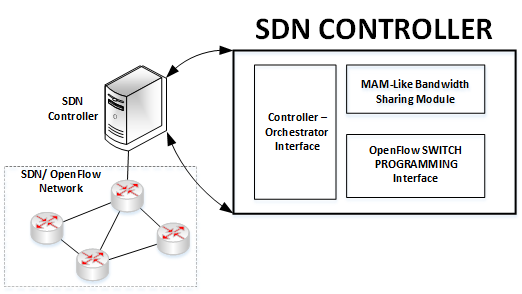}
\caption{ SDN controller modules}
\label{SDNController}
\end{figure}

The SDN controller functional blocks are illustrated in Figure \ref{SDNController}:
\begin{itemize}
    \item   An interface with the PSIoT orchestrator;
    \item   A MAM-like bandwidth sharing module (MAM - Maximum Allocation Model); and
    \item   An SDN/OpenFlow network-programming interface.
\end{itemize}

The SDN controller interface communicates with the PSIoT orchestrator and is responsible for interpreting commands coming from the orchestrator requesting the setup of communication paths between IoT aggregators (producers) and IoT consumers.

The MAM-like module is a bandwidth sharing schema that uses the bandwidth allocation model strategy defined for the MAM model. Bandwidth sharing is an often used concept explored in Bandwidth Allocation Models (BAM) that distributes and efficiently manages scarce network resources \cite{GBAM}. The MAM-like module basically keeps track over the used bandwidth for all links over the path between IoT producer and IoT consumer.

The MAM BAM model assumes that the link bandwidth is divided without any sharing between a set of traffic classes \cite{}. For each traffic class (TC) is allocated a slice of the bandwidth and network users are allocated and allowed to use only the bandwidth allocated for its class. This limit is denominated bandwidth constraint (BC).

In the PSIoT-SDN framework the traffic classes and bandwidth constraints were defined and configured as follows:

\begin{itemize}
    \item   Traffic class 0 (TC0) supports all low priority traffic, including from the IoT, in the network;
    \item   Traffic class 1 (TC1) supports some mid-priority non-IoT traffic; and
    \item   Traffic class 2 (TC2) supports high priority traffic other than IoT over the network.
\end{itemize}

The bandwidth allocated for each traffic class is a management decision configured as BC0, BC1 and BC2.

The SDN controller network-programming interface configures the switches to establish paths for IoT traffic over the network. Paths between producers and consumers are previously calculated using any routing algorithm. The SDN controller uses this computed paths to configure the switch flow-tables over the path.

The SDN controller monitors the network bandwidth usage and network-level IoT class bandwidth allocations. As the traffic changes, the controller can allocate additional bandwidth to IoT traffic by reassigning the TC of IoT traffic flows to TC's with  available bandwidth in the links. This is done by the PSIoT-Orch communicating grouped traffic flows, in the form of origin-destination pairs, to the MAM-like module at the SDN controller. This allows the IoT traffic growing on demand to maintain appropriate throughput, by allowing the PSIoT-SDN to utilize the additional bandwidth allocated by the controller to increase the transmission rates of IoT data, according to their QoS need.

\begin{figure}[htbp]
\centering
\includegraphics[width=\columnwidth]{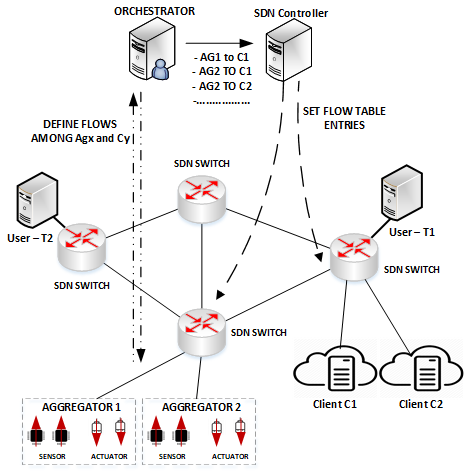}
\caption{ SDN controller operation}
\label{SDNControllerOperation}
\end{figure}

The SDN controller operation flow is illustrated in Figure \ref{SDNControllerOperation}.

The steps involved are the following:
\begin{itemize}
    \item   The orchestrator checks QoS status on the aggregators and defines IoT flows between producer and consumers.
    \item   IoT flows are created upon orchestrator requests to the SDN controller.
    \item   The SDN controller creates entries on the OpenFlow switches involved in the path.
    \item   SDN controller adjusts IoT queue bandwidth limits at the switches in accordance with bandwidth availability controlled by the MAM-like module.
\end{itemize}

The PSIoT-SDN framework assumes that all traffic flowing in the network are queued at the switches and have bandwidth constraints defined. In effect, the queue implementation and corresponding bandwidth limits are a feature that has dependency of the switched and network operating system used. This is a limitation of the proposed framework but is essential to have bandwidth limits to guarantee QoS requirements at network level \cite{krishna_providing_2016-1}.

\section{PSIoT-SDN Proof of Concept} \label{sssec:evaluation}

An implementation of the PSIoT-SDN framework with integrated SDN  is made to be easily tested in a simulated environment.

The PSIoT-SDN proof of concept components are:
\begin{itemize}
    \item   A network topology created with the Mininet emulator \cite{lantz_mininet_2018}; 
    \item   The POX \cite{pox} SDN network operating system supporting the IoT-SDN network management;
    \item   Open vSwitch switches  \cite{pfaff2015design}, controlled by the OpenFlow protocol;
    \item   User traffic generators;
    \item   The MAM-like BAM module implemented with OpenStack tools; and
    \item   The SDN user-controller software.
\end{itemize}




The proof of concept scenario presents a simple and controlled network topology using links with constrained resources. Its topology and network resources are set in such a way as to present both IoT and normal traffic competing for resources.

Figure \ref{SDNControllerOperation} shows the proof of concept topology with the PSIoT-SDN components on the network. The proof of concept traffic configuration adopted is as follows:

\begin{itemize}
    \item   The network link resource allocation is set to 3 traffic classes (TC0, TC1 and TC2)
    \item   TC0 is configured initially with 50\% of link bandwidth (BC0);
    \item   TC1 is configured initially with 30\%  (BC1);
    \item   TC2 is configured initially with 20\%  (BC2);
    \item   IoT traffic is allocated at TC0; and
    \item   All other traffic is allocated at TC1 and TC2 according with their priority.
\end{itemize}

The MiniNet emulated OpenFlow switch (Open vSwicth) uses bandwidth controlled network queues (NQ0, NQ1 and NQ2), each of then supporting a traffic class (TC0, TC1 and TC2) \cite{krishna_providing_2016}. The link's available bandwidth is restrained for  traffic flows belonging to each class level. This division is done to simulate real world network constraints on traffic, according to connection SLA's, and IoT traffic is assigned to the lowest priority class to simulate consumer-level internet connections.

This scenario is designed to match common network configurations and to display the better QoS control that the PSIoT-SDN framework can achieve when integrated into an SDN network's management. This will be accomplished by running an implementation of the PSIoT-SDN, which preserves the PSIoT-Orch queue management control at the aggregators, inside an emulated SDN network in Mininet, along with a custom SDN controller that will communicate with the Orchestrator.

\subsection{Proof of concept simulated runs}

This proof of concept simulation aims to show how integrating with the SDN controller has a positive effect on the PSIoT-SDN's network use and on its ability to provide better QoS guarantees to its users.

\begin{figure}[htbp]
\centering
\includegraphics[width=1.05\columnwidth]{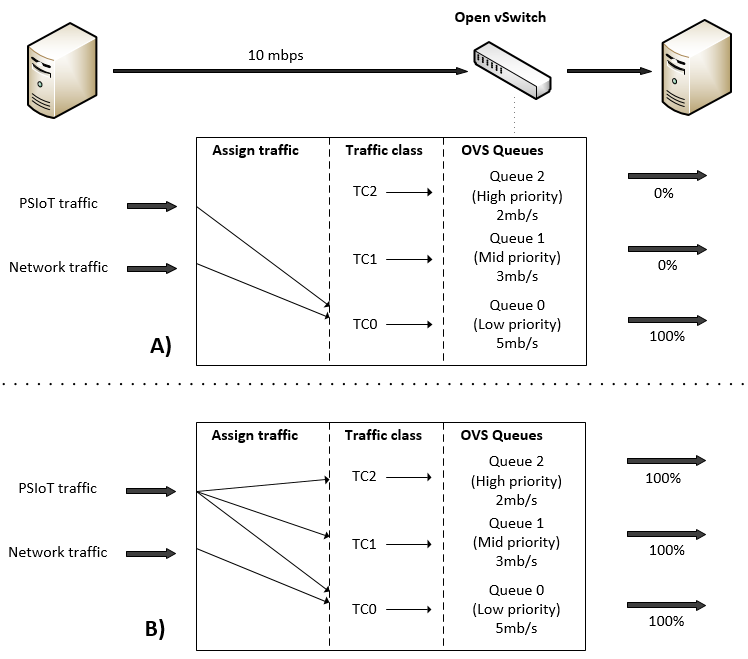}
\caption{SDN controller reassigns traffic to TC's. A) TC assignment before and B) after reassigning PSIoT-SDN traffic }
\label{vswitch-compare}
\end{figure}

To this effect, the SDN integration implements this in a way where PSIoT-SDN traffic groups, as signaled to the SDN controller, can be assigned to TC's with spare capacity. Figure \ref{vswitch-compare} shows how the SDN controller reassigns flows of traffic from the PSIoT-SDN to TC's with available bandwidth.

The above network scenario is simulated with a series of events that aim to show the strengths of SDN-enabled bandwidth sharing with the integrating of the SDN network management to the PSIoT-SDN. 

The events are ordered as follows:

\begin{enumerate}
    \item T1 sends traffic to T2. This traffic is in the TC0 class.
    \item C1 subscribes to Ag1 and Ag2. All IoT traffic is in the TC0 class.
    \item C2 subscribes to Ag1 and Ag2.
    \item T1 stops sending traffic.
    \item T2 sends traffic to T1. This traffic is in the TC1 class.
    \item T2 stops sending traffic.
    \item End of simulation.
\end{enumerate}


\subsection{Evaluating the results} 

BAMs allow better network utilization by sharing unused bandwidth between TC's.  They have a positive effect on network traffic as a whole. However, the advantages gained from BAM's generally don't favor targeted applications in the network. Simply having dynamic BAM's like G-BAM \cite{GBAM} isn't enough to increase the QoS control that the PSIoT-SDN has on its traffic because it still has to compete for the extra bandwidth.

Integration with a custom SDN BAM solution allows the PSIoT-SDN to passively signal to the SDN Controller which traffic groupings that will take precedence in bandwidth sharing, giving the PSIoT-SDN a more exclusive access to the currently unused bandwidth.

With the simulation we can observe how the SDN controller allocates bandwidth to PSIoT-SDN traffic groups at the network and how the framework responds to variations in bandwidth demand. In order to measure and compare the effects of the SDN integration, the proof of concept will run the PSIoT-SDN, that integrates SDN/ OpenFlow, and the PSIoT-Orch that does not make use of SDN.

\begin{figure}[htbp]
\centering
\includegraphics[width=\columnwidth]{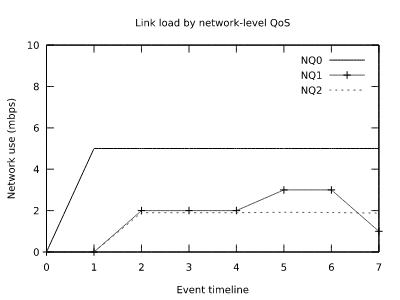}
\caption{ Link load during simulation events }
\label{linkutil}
\end{figure}

Figure \ref{linkutil} shows the link load at each simulation event. By redistributing the traffic from aggregators, according to IoT QoS levels, into different  traffic classes there is a better use of the overall network bandwidth and also provides the IoT traffic with network-level QoS.

\begin{figure}[htbp]
\centering
\includegraphics[width=1.05\columnwidth]{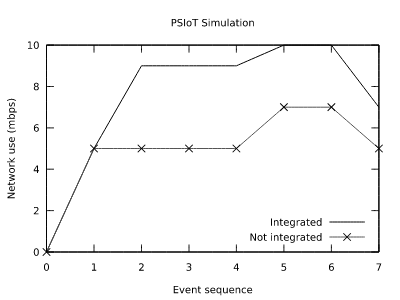}
\caption{ Link bandwidth utilization with integrated \& non-integrated SDN}
\label{linkcomp}
\end{figure}

Comparing the effects of the PSIoT-SDN integration, Figure \ref{linkcomp} shows the link load of the simulation being executed  with and without network-level QoS integration. The integration with the SDN network can utilize up to 100\% of the link's capacity and an average of 86\% bandwidth utilization. When not integrated, the IoT traffic output and overall network utilization is limited by the constraints of the applications associated traffic class.

BAM's are not a novel concept, and different flavours allow for similar bandwidth sharing. However the main advantage of this integration is the ability to share bandwidth according to IoT specific QoS levels, enhancing the level of guarantees in QoS capabilities for IoT traffic. These results display the flexibility of the PSIoT-SDN integrated into the network's management, especially considering how SDN networks have facilitated integration with Application Domain services and more easily allow for third party integration into network management.

The orchestrator implements IoT data QoS by effectively managing transmission rates. The lower the rate of data transmission, the longer the sending of information is delayed. Considering that transmission rates are the determining factor for the QoS enforced on data going trough the PSIoT-SDN framework, an increase in overall network utilization will decrease the overall data transmission delay, providing better QoS.



A relevant consideration concerning the proof of concept topology is that the evaluation considered a single network link. In effect, our objective is to demonstrate that the cumulative edge-QoS and network-QoS control will enhance the overall quality of service for massive IoT traffic. It is also worth to emphasize that BAMs manage bandwidth on a per-link basis \cite{reale_alloctc-sharing_2011-1}. As such, the overall QoS behavior for a path between IoT data producer and consumer will benefit from any link bandwidth improvement over the path. This further validates the proof of concept presented.

\section{Final Considerations} \label{sssec:conclusion}

The internet has changed how people, machines and physical objects, "things", communicate. The use of "things", devices that communicate with each other and with people has made an impact on the current network technology, creating issues with scalability, security and performance.

The solutions to internet quality problems already exist for quite some time. Traffic from specific technologies that have a very large impact on network congestion, like the IoT, can leverage these solutions of routing, geography and bandwidth sharing to better serve it's QoS requirements by creating customized frameworks and protocols.

Factors such as how data is transferred, discovered, shared and consumed are fundamental in creating technology and managing network traffic that better fits the needs of the IoT.

We presented an overview of the requirements and characteristics of IoT data, how devices communicate and how IoT data is transmitted. Based on these attributes, we introduced a framework for IoT traffic orchestration, enabling dynamic QoS capabilities and further integration into the backbone network management, namely SDN.

The framework Orchestrator provides Fog-like IoT data aggregators on the network edge with a familiar Pub/Sub interface for IoT consumers. The IoT data's transmission is managed by IoT specific QoS classes, enforced by data transmission scheduling. 

The framework Orchestrator also orchestrates subscribed traffic from all registered IoT aggregators, according to available information on bandwidth and topology and subscription QoS requirements. The QoS capabilities can further be enhanced by integrating into the SDN management.

Resource management and traffic scheduling are often, and in the scenario explored in this paper, a zero-sum game. The framework Orchestrator manages traffic with the effect of providing QoS to high priority IoT data, as asserted by IoT topic subscriptions. This effect is increased when SDN integration allows network bandwidth sharing to be passively managed by the PSIoT-SDN framework, enabling better utilization of the network and enhanced QoS fulfillment for IoT traffic.

The simulation results showed how the shared bandwidth was distributed among the framework's IoT QoS levels, generating a larger throughput of IoT traffic when compared to the same scenario when SDN integration was excluded. With these positive results, the PSIoT-SDN framework is validated and displays its usefulness in managing QoS for IoT traffic.

In summary, the main contribution of the PSIoT-SDN framework is to couple a QoS approach on network edge with a network bandwidth sharing strategy that is SDN-controlled.

The PSIoT-SDN framework stretches across the network, and connects several different IoT networks and domains. Future work can consider expanding on the SDN integration in the Network Domain, tackling the discoverability of IoT topics in the Pub/Sub aggregator nodes, as well as aggregation and data policies in the M2M domain.

The framework was built to integrate into existing internet technologies, working over public networks or with enhanced capabilities in SDN networks. This is an important characteristic to maintain in an IoT framework, considering the fast changing scenario and the heterogeneity of data, traffic and applications.

\bibliographystyle{ACM-Reference-Format}
\bibliography{ICFNDS}


\begin{thebibliography}{38}


\ifx \showCODEN    \undefined \def \showCODEN     #1{\unskip}     \fi
\ifx \showDOI      \undefined \def \showDOI       #1{#1}\fi
\ifx \showISBNx    \undefined \def \showISBNx     #1{\unskip}     \fi
\ifx \showISBNxiii \undefined \def \showISBNxiii  #1{\unskip}     \fi
\ifx \showISSN     \undefined \def \showISSN      #1{\unskip}     \fi
\ifx \showLCCN     \undefined \def \showLCCN      #1{\unskip}     \fi
\ifx \shownote     \undefined \def \shownote      #1{#1}          \fi
\ifx \showarticletitle \undefined \def \showarticletitle #1{#1}   \fi
\ifx \showURL      \undefined \def \showURL       {\relax}        \fi
\providecommand\bibfield[2]{#2}
\providecommand\bibinfo[2]{#2}
\providecommand\natexlab[1]{#1}
\providecommand\showeprint[2][]{arXiv:#2}

\bibitem[\protect\citeauthoryear{Abu-Elkheir, Hayajneh, and Ali}{Abu-Elkheir
  et~al\mbox{.}}{2013}]%
        {10.3390/s131115582}
\bibfield{author}{\bibinfo{person}{Mervat Abu-Elkheir},
  \bibinfo{person}{Mohammad Hayajneh}, {and} \bibinfo{person}{Najah Ali}.}
  \bibinfo{year}{2013}\natexlab{}.
\newblock \showarticletitle{Data Management for the Internet of Things: Design
  Primitives and Solution}.
\newblock \bibinfo{journal}{\emph{Sensors}} \bibinfo{volume}{13},
  \bibinfo{number}{11} (\bibinfo{date}{nov} \bibinfo{year}{2013}),
  \bibinfo{pages}{15582--15612}.
\newblock
\urldef\tempurl%
\url{https://doi.org/10.3390/s131115582}
\showDOI{\tempurl}


\bibitem[\protect\citeauthoryear{Al-Falahy and Alani}{Al-Falahy and
  Alani}{2017}]%
        {10.1109/CEEC.2017.8101604}
\bibfield{author}{\bibinfo{person}{Naser Al-Falahy} {and} \bibinfo{person}{Omar
  Y.~K. Alani}.} \bibinfo{year}{2017}\natexlab{}.
\newblock \showarticletitle{Supporting massive M2M traffic in the Internet of
  Things using millimetre wave 5G network}. In \bibinfo{booktitle}{\emph{2017
  9th Computer Science and Electronic Engineering ({CEEC})}}.
  \bibinfo{publisher}{{IEEE}}.
\newblock
\urldef\tempurl%
\url{https://doi.org/10.1109/ceec.2017.8101604}
\showDOI{\tempurl}


\bibitem[\protect\citeauthoryear{Al-Fuqaha, Guizani, Mohammadi, Aledhari, and
  Ayyash}{Al-Fuqaha et~al\mbox{.}}{2015}]%
        {10.1109/COMST.2015.2444095}
\bibfield{author}{\bibinfo{person}{Ala Al-Fuqaha}, \bibinfo{person}{Mohsen
  Guizani}, \bibinfo{person}{Mehdi Mohammadi}, \bibinfo{person}{Mohammed
  Aledhari}, {and} \bibinfo{person}{Moussaq Ayyash}.}
  \bibinfo{year}{2015}\natexlab{}.
\newblock \showarticletitle{Internet of Things: A Survey on Enabling
  Technologies, Protocols, and Applications}.
\newblock \bibinfo{journal}{\emph{{IEEE} Communications Surveys {\&}
  Tutorials}} \bibinfo{volume}{17}, \bibinfo{number}{4} (\bibinfo{year}{2015}),
  \bibinfo{pages}{2347--2376}.
\newblock
\urldef\tempurl%
\url{https://doi.org/10.1109/comst.2015.2444095}
\showDOI{\tempurl}


\bibitem[\protect\citeauthoryear{Al-Shammari, Al-Aboody, and
  Al-Raweshidy}{Al-Shammari et~al\mbox{.}}{2018}]%
        {10.1109/JIOT.2017.2785219}
\bibfield{author}{\bibinfo{person}{Basim K.~J. Al-Shammari},
  \bibinfo{person}{Nadia Al-Aboody}, {and} \bibinfo{person}{Hamed~S.
  Al-Raweshidy}.} \bibinfo{year}{2018}\natexlab{}.
\newblock \showarticletitle{{IoT} Traffic Management and Integration in the
  {QoS} Supported Network}.
\newblock \bibinfo{journal}{\emph{{IEEE} Internet of Things Journal}}
  \bibinfo{volume}{5}, \bibinfo{number}{1} (\bibinfo{date}{feb}
  \bibinfo{year}{2018}), \bibinfo{pages}{352--370}.
\newblock
\urldef\tempurl%
\url{https://doi.org/10.1109/jiot.2017.2785219}
\showDOI{\tempurl}


\bibitem[\protect\citeauthoryear{Alturki, Reiff-Marganiec, and Perera}{Alturki
  et~al\mbox{.}}{2017}]%
        {alturki2017hybrid}
\bibfield{author}{\bibinfo{person}{Badraddin Alturki}, \bibinfo{person}{Stephan
  Reiff-Marganiec}, {and} \bibinfo{person}{Charith Perera}.}
  \bibinfo{year}{2017}\natexlab{}.
\newblock \showarticletitle{A hybrid approach for data analytics for internet
  of things}. In \bibinfo{booktitle}{\emph{Proceedings of the Seventh
  International Conference on the Internet of Things}}. ACM,
  \bibinfo{pages}{7}.
\newblock


\bibitem[\protect\citeauthoryear{Baddeley, Nejabati, Oikonomou, Sooriyabandara,
  and Simeonidou}{Baddeley et~al\mbox{.}}{2018}]%
        {baddeley_evolving_2018}
\bibfield{author}{\bibinfo{person}{Michael Baddeley}, \bibinfo{person}{Reza
  Nejabati}, \bibinfo{person}{George Oikonomou}, \bibinfo{person}{Mahesh
  Sooriyabandara}, {and} \bibinfo{person}{Dimitra Simeonidou}.}
  \bibinfo{year}{2018}\natexlab{}.
\newblock \showarticletitle{Evolving {{SDN}} for {{Low}}-{{Power IoT
  Networks}}}. In \bibinfo{booktitle}{\emph{2018 4th {{IEEE Conference}} on
  {{Network Softwarization}} and {{Workshops}} ({{NetSoft}})}}.
  \bibinfo{pages}{71--79}.
\newblock
\urldef\tempurl%
\url{https://doi.org/10.1109/NETSOFT.2018.8460125}
\showDOI{\tempurl}


\bibitem[\protect\citeauthoryear{Bera, Misra, and Vasilakos}{Bera
  et~al\mbox{.}}{2017}]%
        {bera_software-defined_2017-1}
\bibfield{author}{\bibinfo{person}{Samaresh Bera}, \bibinfo{person}{Sudip
  Misra}, {and} \bibinfo{person}{Athanasios~V. Vasilakos}.}
  \bibinfo{year}{2017}\natexlab{}.
\newblock \showarticletitle{Software-{{Defined Networking}} for {{Internet}} of
  {{Things}}: {{A Survey}}}.
\newblock \bibinfo{journal}{\emph{IEEE Internet of Things Journal}}
  \bibinfo{volume}{4}, \bibinfo{number}{6} (\bibinfo{date}{Dec.}
  \bibinfo{year}{2017}), \bibinfo{pages}{1994--2008}.
\newblock
\showISSN{2327-4662}
\urldef\tempurl%
\url{https://doi.org/10.1109/JIOT.2017.2746186}
\showDOI{\tempurl}


\bibitem[\protect\citeauthoryear{Bonomi, Milito, Zhu, and Addepalli}{Bonomi
  et~al\mbox{.}}{2012}]%
        {bonomi2012fog}
\bibfield{author}{\bibinfo{person}{Flavio Bonomi}, \bibinfo{person}{Rodolfo
  Milito}, \bibinfo{person}{Jiang Zhu}, {and} \bibinfo{person}{Sateesh
  Addepalli}.} \bibinfo{year}{2012}\natexlab{}.
\newblock \showarticletitle{Fog computing and its role in the internet of
  things}. In \bibinfo{booktitle}{\emph{Proceedings of the first edition of the
  MCC workshop on Mobile cloud computing}}. ACM, \bibinfo{pages}{13--16}.
\newblock


\bibitem[\protect\citeauthoryear{Borgia}{Borgia}{2014a}]%
        {borgia2014}
\bibfield{author}{\bibinfo{person}{Eleonora Borgia}.}
  \bibinfo{year}{2014}\natexlab{a}.
\newblock \showarticletitle{The Internet of Things vision: Key features,
  applications and open issues}.
\newblock \bibinfo{journal}{\emph{Computer Communications}}
  \bibinfo{volume}{54} (\bibinfo{year}{2014}), \bibinfo{pages}{1--31}.
\newblock
\urldef\tempurl%
\url{https://doi.org/10.1016/j.comcom.2014.09.008}
\showDOI{\tempurl}


\bibitem[\protect\citeauthoryear{Borgia}{Borgia}{2014b}]%
        {10.1016/j.comcom.2014.09.008}
\bibfield{author}{\bibinfo{person}{Eleonora Borgia}.}
  \bibinfo{year}{2014}\natexlab{b}.
\newblock \showarticletitle{The Internet of Things vision: Key features,
  applications and open issues}.
\newblock \bibinfo{journal}{\emph{Computer Communications}}
  \bibinfo{volume}{54} (\bibinfo{date}{dec} \bibinfo{year}{2014}),
  \bibinfo{pages}{1--31}.
\newblock
\urldef\tempurl%
\url{https://doi.org/10.1016/j.comcom.2014.09.008}
\showDOI{\tempurl}


\bibitem[\protect\citeauthoryear{Boussard, Bui, Ciavaglia, Douville, Pallec,
  Sauze, Noirie, Papillon, Peloso, and Santoro}{Boussard et~al\mbox{.}}{2015}]%
        {boussard_software-defined_2015-1}
\bibfield{author}{\bibinfo{person}{Mathieu Boussard}, \bibinfo{person}{Dinh~T.
  Bui}, \bibinfo{person}{Laurent Ciavaglia}, \bibinfo{person}{Richard
  Douville}, \bibinfo{person}{Michel~Le Pallec}, \bibinfo{person}{Nicolas~Le
  Sauze}, \bibinfo{person}{Ludovic Noirie}, \bibinfo{person}{Serge Papillon},
  \bibinfo{person}{Pierre Peloso}, {and} \bibinfo{person}{Francesco Santoro}.}
  \bibinfo{year}{2015}\natexlab{}.
\newblock \showarticletitle{Software-{{Defined LANs}} for {{Interconnected
  Smart Environment}}}.
\newblock  (\bibinfo{date}{Sept.} \bibinfo{year}{2015}).
\newblock


\bibitem[\protect\citeauthoryear{Institute}{Institute}{2012}]%
        {etsi}
\bibfield{author}{\bibinfo{person}{ETSI European Telecommunications~Standards
  Institute}.} \bibinfo{year}{2012}\natexlab{}.
\newblock \bibinfo{title}{Machine-to-Machine Communications (M2M); mIa, dIa and
  mId interfaces}.
\newblock
\newblock
\urldef\tempurl%
\url{http://www.etsi.org/deliver/etsi_ts/102900_102999/102921/01.01.01_60/ts_102921v010101p.pdf}
\showURL{%
\tempurl}


\bibitem[\protect\citeauthoryear{Jararweh, {Al-Ayyoub}, Darabseh, Benkhelifa,
  Vouk, and Rindos}{Jararweh et~al\mbox{.}}{2015}]%
        {jararweh_sdiot_2015}
\bibfield{author}{\bibinfo{person}{Yaser Jararweh}, \bibinfo{person}{Mahmoud
  {Al-Ayyoub}}, \bibinfo{person}{Ala' Darabseh}, \bibinfo{person}{Elhadj
  Benkhelifa}, \bibinfo{person}{Mladen Vouk}, {and} \bibinfo{person}{Andy
  Rindos}.} \bibinfo{year}{2015}\natexlab{}.
\newblock \showarticletitle{{{SDIoT}}: {{A Software Defined Based Internet}} of
  {{Things Framework}}}.
\newblock \bibinfo{journal}{\emph{Journal of Ambient Intelligence and Humanized
  Computing}} \bibinfo{volume}{6}, \bibinfo{number}{4} (\bibinfo{date}{Aug.}
  \bibinfo{year}{2015}), \bibinfo{pages}{453--461}.
\newblock
\showISSN{1868-5145}
\urldef\tempurl%
\url{https://doi.org/10.1007/s12652-015-0290-y}
\showDOI{\tempurl}


\bibitem[\protect\citeauthoryear{Jiang, Huang, and Tsang}{Jiang
  et~al\mbox{.}}{2017}]%
        {10.1109/MNET.2017.1700271}
\bibfield{author}{\bibinfo{person}{Yuxuan Jiang}, \bibinfo{person}{Zhe Huang},
  {and} \bibinfo{person}{Danny H.~K. Tsang}.} \bibinfo{year}{2017}\natexlab{}.
\newblock \showarticletitle{Challenges and Solutions in Fog Computing
  Orchestration}.
\newblock \bibinfo{journal}{\emph{{IEEE} Network}} (\bibinfo{year}{2017}),
  \bibinfo{pages}{1--8}.
\newblock
\urldef\tempurl%
\url{https://doi.org/10.1109/mnet.2017.1700271}
\showDOI{\tempurl}


\bibitem[\protect\citeauthoryear{Kreutz, Ramos, Verissimo, Rothenberg,
  Azodolmolky, and Uhlig}{Kreutz et~al\mbox{.}}{2015}]%
        {kreutz2015software}
\bibfield{author}{\bibinfo{person}{Diego Kreutz}, \bibinfo{person}{Fernando~MV
  Ramos}, \bibinfo{person}{Paulo~Esteves Verissimo},
  \bibinfo{person}{Christian~Esteve Rothenberg}, \bibinfo{person}{Siamak
  Azodolmolky}, {and} \bibinfo{person}{Steve Uhlig}.}
  \bibinfo{year}{2015}\natexlab{}.
\newblock \showarticletitle{Software-defined networking: A comprehensive
  survey}.
\newblock \bibinfo{journal}{\emph{Proc. IEEE}} \bibinfo{volume}{103},
  \bibinfo{number}{1} (\bibinfo{year}{2015}), \bibinfo{pages}{14--76}.
\newblock


\bibitem[\protect\citeauthoryear{Krishna}{Krishna}{2016}]%
        {krishna_providing_2016-1}
\bibfield{author}{\bibinfo{person}{Hedi Krishna}.}
  \bibinfo{year}{2016}\natexlab{}.
\newblock \emph{\bibinfo{title}{Providing {{End}}-to-End {{Bandwidth
  Guarantees}} with {{OpenFlow}}}}.
\newblock \bibinfo{thesistype}{Ph.D. Dissertation}.
\newblock


\bibitem[\protect\citeauthoryear{Krishna, van Adrichem, and Kuipers}{Krishna
  et~al\mbox{.}}{2016}]%
        {krishna_providing_2016}
\bibfield{author}{\bibinfo{person}{Hedi Krishna}, \bibinfo{person}{Niels L.~M.
  van Adrichem}, {and} \bibinfo{person}{Fernando~A. Kuipers}.}
  \bibinfo{year}{2016}\natexlab{}.
\newblock \showarticletitle{Providing {{Bandwidth Guarantees}} with
  {{Openflow}}}. In \bibinfo{booktitle}{\emph{2016 {{Symposium}} on
  {{Communications}} and {{Vehicular Technologies}} ({{SCVT}})}}.
  \bibinfo{pages}{1--6}.
\newblock
\urldef\tempurl%
\url{https://doi.org/10.1109/SCVT.2016.7797664}
\showDOI{\tempurl}


\bibitem[\protect\citeauthoryear{Lantz}{Lantz}{2018}]%
        {lantz_mininet_2018}
\bibfield{author}{\bibinfo{person}{Bob Lantz}.}
  \bibinfo{year}{2018}\natexlab{}.
\newblock \bibinfo{title}{Mininet: {{Emulator}} for {{Rapid Prototyping}} of
  {{Software Defined Networks}}}.
\newblock
\newblock


\bibitem[\protect\citeauthoryear{Mc and Scott}{Mc and Scott}{2013}]%
        {pox}
\bibfield{author}{\bibinfo{person}{Murphy Mc} {and} \bibinfo{person}{Colin
  Scott}.} \bibinfo{year}{2013}\natexlab{}.
\newblock \bibinfo{title}{POX is a networking software platform written in
  Python}.
\newblock
\newblock


\bibitem[\protect\citeauthoryear{Modarresi, Gangadhar, and Sterbenz}{Modarresi
  et~al\mbox{.}}{2017}]%
        {10.1109/RNDM.2017.8093036}
\bibfield{author}{\bibinfo{person}{Amir Modarresi}, \bibinfo{person}{Siddharth
  Gangadhar}, {and} \bibinfo{person}{James~P.G. Sterbenz}.}
  \bibinfo{year}{2017}\natexlab{}.
\newblock \showarticletitle{A framework for improving network resilience using
  {SDN} and fog nodes}. In \bibinfo{booktitle}{\emph{2017 9th International
  Workshop on Resilient Networks Design and Modeling ({RNDM})}}.
  \bibinfo{publisher}{{IEEE}}.
\newblock
\urldef\tempurl%
\url{https://doi.org/10.1109/rndm.2017.8093036}
\showDOI{\tempurl}


\bibitem[\protect\citeauthoryear{Moraes, Reale, and Martins}{Moraes
  et~al\mbox{.}}{2018}]%
        {moraes_publish/subscribe_2018-2}
\bibfield{author}{\bibinfo{person}{Pedro~F. Moraes}, \bibinfo{person}{Rafael~F.
  Reale}, {and} \bibinfo{person}{Joberto S.~B. Martins}.}
  \bibinfo{year}{2018}\natexlab{}.
\newblock \showarticletitle{A {{Publish}}/{{Subscribe QoS}}-Aware {{Framework}}
  for {{Massive IoT Traffic Orchestration}}}. In
  \bibinfo{booktitle}{\emph{Proceedings of the 6th {{International Workshop}}
  on {{ADVANCEs}} in {{ICT Infrastructures}} and {{Services}} - {{ADVANCE}}
  2018}}. \bibinfo{address}{Santiago, Chile}, \bibinfo{pages}{1--14}.
\newblock
\urldef\tempurl%
\url{https://doi.org/10.5281/zenodo.1098298}
\showDOI{\tempurl}


\bibitem[\protect\citeauthoryear{Pfaff, Pettit, Koponen, Jackson, Zhou,
  Rajahalme, Gross, Wang, Stringer, Shelar, et~al\mbox{.}}{Pfaff
  et~al\mbox{.}}{2015}]%
        {pfaff2015design}
\bibfield{author}{\bibinfo{person}{Ben Pfaff}, \bibinfo{person}{Justin Pettit},
  \bibinfo{person}{Teemu Koponen}, \bibinfo{person}{Ethan Jackson},
  \bibinfo{person}{Andy Zhou}, \bibinfo{person}{Jarno Rajahalme},
  \bibinfo{person}{Jesse Gross}, \bibinfo{person}{Alex Wang},
  \bibinfo{person}{Joe Stringer}, \bibinfo{person}{Pravin Shelar},
  {et~al\mbox{.}}} \bibinfo{year}{2015}\natexlab{}.
\newblock \showarticletitle{The design and implementation of open vswitch}. In
  \bibinfo{booktitle}{\emph{12th $\{$USENIX$\}$ Symposium on Networked Systems
  Design and Implementation ($\{$NSDI$\}$ 15)}}. \bibinfo{pages}{117--130}.
\newblock


\bibitem[\protect\citeauthoryear{Pourghebleh and Navimipour}{Pourghebleh and
  Navimipour}{2017}]%
        {10.1016/j.jnca.2017.08.006}
\bibfield{author}{\bibinfo{person}{Behrouz Pourghebleh} {and}
  \bibinfo{person}{Nima~J. Navimipour}.} \bibinfo{year}{2017}\natexlab{}.
\newblock \showarticletitle{Data Aggregation Mechanisms in the Internet of
  Things: A Systematic Review of the Literature and Recommendations for Future
  Research}.
\newblock \bibinfo{journal}{\emph{Journal of Network and Comp. Applications}}
  \bibinfo{volume}{97} (\bibinfo{date}{nov} \bibinfo{year}{2017}),
  \bibinfo{pages}{23--34}.
\newblock
\urldef\tempurl%
\url{https://doi.org/10.1016/j.jnca.2017.08.006}
\showDOI{\tempurl}


\bibitem[\protect\citeauthoryear{Qian, Luo, Du, and Guo}{Qian
  et~al\mbox{.}}{2009}]%
        {cloud}
\bibfield{author}{\bibinfo{person}{Ling Qian}, \bibinfo{person}{Zhiguo Luo},
  \bibinfo{person}{Yujian Du}, {and} \bibinfo{person}{Leitao Guo}.}
  \bibinfo{year}{2009}\natexlab{}.
\newblock \showarticletitle{Cloud Computing: An Overview}.
\newblock \bibinfo{journal}{\emph{Lecture Notes in Computer Science}}
  (\bibinfo{year}{2009}), \bibinfo{pages}{626--631}.
\newblock
\urldef\tempurl%
\url{https://doi.org/10.1007/978-3-642-10665-1_63}
\showDOI{\tempurl}


\bibitem[\protect\citeauthoryear{Rachikidi}{Rachikidi}{2017}]%
        {These-Elie}
\bibfield{author}{\bibinfo{person}{Elie~E. Rachikidi}.}
  \bibinfo{year}{2017}\natexlab{}.
\newblock \bibinfo{title}{Modeling and Placement Optimization of Compound
  Service in a Converged Infrastructure of Cloud Computing and Internet of
  Things}.  (\bibinfo{year}{2017}).
\newblock
\newblock
\shownote{Universite Paris-Saclay, Evry.}


\bibitem[\protect\citeauthoryear{Rahman, Ahmed, and Hussain}{Rahman
  et~al\mbox{.}}{2016}]%
        {10.1109/WiSPNET.2016.7566346}
\bibfield{author}{\bibinfo{person}{Hafizur Rahman}, \bibinfo{person}{Nurzaman
  Ahmed}, {and} \bibinfo{person}{Iftekhar Hussain}.}
  \bibinfo{year}{2016}\natexlab{}.
\newblock \showarticletitle{Comparison of Data Aggregation Techniques in
  Internet of Things ({IoT})}. In \bibinfo{booktitle}{\emph{2016 International
  Conference on Wireless Communications, Signal Processing and Networking
  ({WiSPNET})}}. \bibinfo{publisher}{{IEEE}}.
\newblock
\urldef\tempurl%
\url{https://doi.org/10.1109/wispnet.2016.7566346}
\showDOI{\tempurl}


\bibitem[\protect\citeauthoryear{Reale, Bezerra, and Martins}{Reale
  et~al\mbox{.}}{2014}]%
        {GBAM}
\bibfield{author}{\bibinfo{person}{Rafael~F. Reale},
  \bibinfo{person}{Romildo~M. Bezerra}, {and} \bibinfo{person}{Joberto S.~B.
  Martins}.} \bibinfo{year}{2014}\natexlab{}.
\newblock \showarticletitle{G-BAM: A Generalized Bandwidth Allocation Model for
  IP/MPLS/DS-TE Networks}.
\newblock \bibinfo{journal}{\emph{Inter. Journal of Computer Information
  Systems and Industrial Management Applications}}  \bibinfo{volume}{6}
  (\bibinfo{year}{2014}), \bibinfo{pages}{635--643}.
\newblock
\showISSN{2150-7988}
\urldef\tempurl%
\url{https://doi.org/10.5281/zenodo.1292771}
\showDOI{\tempurl}


\bibitem[\protect\citeauthoryear{Reale, Neto, and Martins}{Reale
  et~al\mbox{.}}{2011}]%
        {reale_alloctc-sharing_2011-1}
\bibfield{author}{\bibinfo{person}{Rafael~F. Reale}, \bibinfo{person}{Walter d
  C.~P. Neto}, {and} \bibinfo{person}{Joberto S.~B. Martins}.}
  \bibinfo{year}{2011}\natexlab{}.
\newblock \showarticletitle{{{AllocTC}}-{{Sharing}}: {{A New Bandwidth
  Allocation Model}} for {{Ds}}-{{Te Networks}}}. In
  \bibinfo{booktitle}{\emph{7th {{Latin American Network Operations}} and
  {{Management Symposium}}}}. \bibinfo{publisher}{{IEEE Institute of Electrical
  and Electronics Engineers}}, \bibinfo{address}{Quito, Equador},
  \bibinfo{pages}{1--4}.
\newblock
\urldef\tempurl%
\url{https://doi.org/10.1109/LANOMS.2011.6102265}
\showDOI{\tempurl}


\bibitem[\protect\citeauthoryear{Santoro, Zozin, Pizzolli, Pellegrini, and
  Cretti}{Santoro et~al\mbox{.}}{2017}]%
        {10.1109/CloudCom.2017.62}
\bibfield{author}{\bibinfo{person}{Daniele Santoro}, \bibinfo{person}{Daniel
  Zozin}, \bibinfo{person}{Daniele Pizzolli}, \bibinfo{person}{Francesco~De
  Pellegrini}, {and} \bibinfo{person}{Silvio Cretti}.}
  \bibinfo{year}{2017}\natexlab{}.
\newblock \showarticletitle{Foggy: A Platform for Workload Orchestration in a
  Fog Computing Environment}. In \bibinfo{booktitle}{\emph{2017 {IEEE}
  International Conference on Cloud Computing Technology and Science
  ({CloudCom})}}. \bibinfo{publisher}{{IEEE}}.
\newblock
\urldef\tempurl%
\url{https://doi.org/10.1109/cloudcom.2017.62}
\showDOI{\tempurl}


\bibitem[\protect\citeauthoryear{Stojmenovic and Wen}{Stojmenovic and
  Wen}{2014}]%
        {fog}
\bibfield{author}{\bibinfo{person}{Ivan Stojmenovic} {and}
  \bibinfo{person}{Sheng Wen}.} \bibinfo{year}{2014}\natexlab{}.
\newblock \showarticletitle{The Fog Computing Paradigm: Scenarios and Security
  Issues}.
\newblock \bibinfo{journal}{\emph{Proceedings of the 2014 Federated Conference
  on Computer Science and Information Systems}} (\bibinfo{year}{2014}).
\newblock
\urldef\tempurl%
\url{https://doi.org/10.15439/2014f503}
\showDOI{\tempurl}


\bibitem[\protect\citeauthoryear{Taneja}{Taneja}{2016}]%
        {10.1109/IC3I.2016.7917982}
\bibfield{author}{\bibinfo{person}{Mukesh Taneja}.}
  \bibinfo{year}{2016}\natexlab{}.
\newblock \showarticletitle{A Framework for Traffic Management in {IoT}
  Networks}. In \bibinfo{booktitle}{\emph{2nd Int. Conference on Contemporary
  Computing and Informatics ({IC}3I)}}.
\newblock
\urldef\tempurl%
\url{https://doi.org/10.1109/ic3i.2016.7917982}
\showDOI{\tempurl}


\bibitem[\protect\citeauthoryear{Tayyaba, Shah, Khan, and Ahmed}{Tayyaba
  et~al\mbox{.}}{2017}]%
        {tayyaba_software_2017}
\bibfield{author}{\bibinfo{person}{Sahrish~Khan Tayyaba},
  \bibinfo{person}{Munam~Ali Shah}, \bibinfo{person}{Omair~Ahmad Khan}, {and}
  \bibinfo{person}{Abdul~Wahab Ahmed}.} \bibinfo{year}{2017}\natexlab{}.
\newblock \showarticletitle{Software {{Defined Network}} ({{SDN}}) {{Based
  Internet}} of {{Things}} ({{IoT}}): {{A Road Ahead}}}. In
  \bibinfo{booktitle}{\emph{Proceedings of the 2nd {{International Conference}}
  on {{Future Networks}} and {{Distributed Systems}}}}
  \emph{(\bibinfo{series}{{{ICFNDS}} '17})}. \bibinfo{publisher}{{ACM}},
  \bibinfo{address}{New York, NY, USA}, \bibinfo{pages}{15:1--15:8}.
\newblock
\showISBNx{978-1-4503-4844-7}
\urldef\tempurl%
\url{https://doi.org/10.1145/3102304.3102319}
\showDOI{\tempurl}


\bibitem[\protect\citeauthoryear{Tripathi, Gupta, and Chourasiya}{Tripathi
  et~al\mbox{.}}{2014}]%
        {tripathi2014survey}
\bibfield{author}{\bibinfo{person}{Ankit Tripathi}, \bibinfo{person}{Sanjeev
  Gupta}, {and} \bibinfo{person}{Bharti Chourasiya}.}
  \bibinfo{year}{2014}\natexlab{}.
\newblock \showarticletitle{Survey on data aggregation techniques for wireless
  sensor networks}.
\newblock \bibinfo{journal}{\emph{International Journal of Advanced Research in
  Computer and Communication Engineering}} \bibinfo{volume}{3},
  \bibinfo{number}{7} (\bibinfo{year}{2014}), \bibinfo{pages}{7366--7371}.
\newblock


\bibitem[\protect\citeauthoryear{Valdivieso~Caraguay, Benito~Peral,
  Barona~Lopez, and Garcia~Villalba}{Valdivieso~Caraguay et~al\mbox{.}}{2014}]%
        {valdivieso2014sdn}
\bibfield{author}{\bibinfo{person}{Angel~Leonardo Valdivieso~Caraguay},
  \bibinfo{person}{Alberto Benito~Peral}, \bibinfo{person}{Lorena~Isabel
  Barona~Lopez}, {and} \bibinfo{person}{Luis~Javier Garcia~Villalba}.}
  \bibinfo{year}{2014}\natexlab{}.
\newblock \showarticletitle{SDN: Evolution and Opportunities in the Development
  IoT Applications}.
\newblock \bibinfo{journal}{\emph{International Journal of Distributed Sensor
  Networks}} \bibinfo{volume}{10}, \bibinfo{number}{5} (\bibinfo{year}{2014}).
\newblock
\urldef\tempurl%
\url{https://doi.org/10.1155/2014/735142}
\showDOI{\tempurl}


\bibitem[\protect\citeauthoryear{Wang, Perera, Jayaraman, Zhang, Strazdins, and
  Ranjan}{Wang et~al\mbox{.}}{2015}]%
        {1506.09118}
\bibfield{author}{\bibinfo{person}{Meisong Wang}, \bibinfo{person}{Charith
  Perera}, \bibinfo{person}{Prem~P. Jayaraman}, \bibinfo{person}{Miranda
  Zhang}, \bibinfo{person}{Peter~E. Strazdins}, {and} \bibinfo{person}{Rajiv
  Ranjan}.} \bibinfo{year}{2015}\natexlab{}.
\newblock \showarticletitle{City Data Fusion: Sensor Data Fusion in the
  Internet of Things}.
\newblock \bibinfo{journal}{\emph{CoRR}}  \bibinfo{volume}{abs/1506.09118}
  (\bibinfo{year}{2015}).
\newblock
\showeprint[arxiv]{1506.09118}
\urldef\tempurl%
\url{http://arxiv.org/abs/1506.09118}
\showURL{%
\tempurl}


\bibitem[\protect\citeauthoryear{Wang, Zhang, and Chen}{Wang
  et~al\mbox{.}}{2018}]%
        {10.1109/CC.2018.8290809}
\bibfield{author}{\bibinfo{person}{Yali Wang}, \bibinfo{person}{Yang Zhang},
  {and} \bibinfo{person}{Junliang Chen}.} \bibinfo{year}{2018}\natexlab{}.
\newblock \showarticletitle{An {SDN}-based Publish/Subscribe-enabled
  Communication Platform for {IoT} Services}.
\newblock \bibinfo{journal}{\emph{China Communications}} \bibinfo{volume}{15},
  \bibinfo{number}{1} (\bibinfo{date}{jan} \bibinfo{year}{2018}),
  \bibinfo{pages}{95--106}.
\newblock
\urldef\tempurl%
\url{https://doi.org/10.1109/cc.2018.8290809}
\showDOI{\tempurl}


\bibitem[\protect\citeauthoryear{Wu, Arkhipov, Asmare, Qin, and McCann}{Wu
  et~al\mbox{.}}{2015}]%
        {wu_ubiflow_2015}
\bibfield{author}{\bibinfo{person}{Di Wu}, \bibinfo{person}{Dmitri~I.
  Arkhipov}, \bibinfo{person}{Eskindir Asmare}, \bibinfo{person}{Zhijing Qin},
  {and} \bibinfo{person}{Julie~A. McCann}.} \bibinfo{year}{2015}\natexlab{}.
\newblock \showarticletitle{Ubiflow: {{Mobility Management}} in
  {{Urban}}-{{Scale Software Defined Iot}}}. In \bibinfo{booktitle}{\emph{2015
  {{IEEE Conference}} on {{Computer Communications}} ({{INFOCOM}})}}.
  \bibinfo{pages}{208--216}.
\newblock
\urldef\tempurl%
\url{https://doi.org/10.1109/INFOCOM.2015.7218384}
\showDOI{\tempurl}


\bibitem[\protect\citeauthoryear{Yu, Liang, He, Hatcher, Lu, Lin, and Yang}{Yu
  et~al\mbox{.}}{2018}]%
        {10.1109/ACCESS.2017.2778504}
\bibfield{author}{\bibinfo{person}{Wei Yu}, \bibinfo{person}{Fan Liang},
  \bibinfo{person}{Xiaofei He}, \bibinfo{person}{William~G. Hatcher},
  \bibinfo{person}{Chao Lu}, \bibinfo{person}{Jie Lin}, {and}
  \bibinfo{person}{Xinyu Yang}.} \bibinfo{year}{2018}\natexlab{}.
\newblock \showarticletitle{A Survey on the Edge Computing for the Internet of
  Things}.
\newblock \bibinfo{journal}{\emph{{IEEE} Access}}  \bibinfo{volume}{6}
  (\bibinfo{year}{2018}), \bibinfo{pages}{6900--6919}.
\newblock
\urldef\tempurl%
\url{https://doi.org/10.1109/access.2017.2778504}
\showDOI{\tempurl}


\end{thebibliography}
\appendix

\end{document}